\documentclass[prl,aps,twocolumn,showpacs]{revtex4}
\usepackage{epsfig}
\include{graphics}
\begin{document}
\title{Scaling and self-similarity in an unforced flow of inviscid fluid
    trapped inside a viscous fluid in a Hele-Shaw cell}
\author{Arkady Vilenkin$^{1}$, Baruch Meerson$^{1}$, and Pavel V. Sasorov$^{2}$}
\affiliation{$^{1}$Racah Institute  of  Physics, Hebrew University
of Jerusalem, Jerusalem 91904, Israel}
\affiliation{$^{2}$Institute of Theoretical and Experimental
Physics, Moscow 117218, Russia}
\begin{abstract}
We investigate quasi-two-dimensional relaxation, by surface tension,
of a long straight stripe of inviscid fluid trapped inside a viscous
fluid in a Hele-Shaw cell.  Combining analytical and numerical
solutions, we describe the emergence of a self-similar dumbbell
shape and find non-trivial dynamic exponents that characterize
scaling behavior of the dumbbell dimensions.
\end{abstract}
 \maketitle

\textit{Introduction.} Consider a bubble of low-viscosity fluid
(say, water) trapped inside a high-viscosity fluid (say, oil) in a
quasi-two-dimensional Hele-Shaw cell. What will happen to the shape
of the bubble, if the (horizontal) plates are perfectly smooth, and
the fluids are immiscible? The answer depends on the initial bubble
shape. A perfectly circular bubble (or an infinite straight stripe)
will not change, while a bubble of any other shape will undergo
surface-tension-driven relaxation until it either becomes a perfect
circle, or breaks into two or more bubbles, which then become
perfect circles. The bubble shape relaxation is non-local, as it is
mediated by a viscous flow in the outer fluid. The resulting free
boundary problem is hard for analysis. This is especially true when
the bubble has a complex (even fractal) shape, like that observed,
in radial geometry, in a \textit{strongly forced} Hele-Shaw flow,
when the viscous fluid was initially displaced by the inviscid fluid
\cite{Sharon}. The shape complexity results from the viscous
fingering instability \cite{ST,Paterson}. The forced Hele-Shaw flow
is a celebrated problem in fluid dynamics and nonlinear dynamics
\cite{Langer1,Kadanoff,Kessler,Casademunt}. The role of small
surface tension there is to introduce a (nontrivial) regularization
on small scales. This Letter deals with an \textit{unforced}
Hele-Shaw (UHS) problem, where surface tension is the only driving
mechanism. We address the UHS problem in the case when the inviscid
fluid is initially in the form of a long stripe. We show that this
special initial condition provides a useful characterization of the
UHS model, as the evolving stripe, which develops a dumbbell shape,
exhibits self-similarity with non-trivial dynamic exponents.

\textit{UHS problem.} Let the inner fluid have negligible
viscosity, so that the pressure inside the bubble is homogeneous.
The velocity of the viscous outer fluid is
$\mathbf{v}\,(\mathbf{r},t)=-(b^2/12\mu) \,\nabla
p\,(\mathbf{r},t)$, where $p$ is the pressure, $\mu$ is the
dynamic viscosity, and $b$ is the plate spacing
\cite{ST,Paterson,Langer1,Kadanoff}. Therefore, the interface
speed is
\begin{equation}\label{speed}
  v_n = - (b^2/12 \mu) \nabla_n p\,,
\end{equation}
where index $n$ denotes the components of the vectors normal to
the interface, and $\nabla_n p$ is evaluated at the respective
points of the interface $\gamma$. In view of incompressibility of
the outer fluid, the pressure is a harmonic function:
\begin{equation}\label{Laplace}
  \nabla^2 p =0\,.
\end{equation}
The Gibbs-Thomson relation at the interface yields
\begin{equation}\label{jump2}
  p\,|_{\gamma} = (\pi/4)\,\sigma {\cal K}\,,
\end{equation}
where $\sigma$ is surface tension, and ${\cal K}$ is the local
curvature of the interface, positive when the inviscid region is
convex outwards. As both the supply of the inner fluid, and
evacuation of the outer fluid are blocked, we demand
\begin{equation}\label{external}
\nabla_n p\,|_{\Gamma} = 0
\end{equation}
at the external boundary of the system $\Gamma$. Equations
(\ref{speed})-(\ref{external}) define the exterior UHS problem
(see Ref. \cite{CLM} for a more detailed discussion). A related,
but different \textit{interior} problem has been also considered,
mainly in the context of singularity formation (pinch-offs) in
bubbles of viscous fluid \cite{Almgren}. The UHS model has two
important properties: (i) the bubble area remains constant, (ii)
the length of the interface between the two fluids is a
non-increasing function of time \cite{Constantin1}.

The UHS problem is not integrable. Moreover, we are unaware of any
analytical solutions to this problem, except for a linear analysis
of a slightly deformed flat or circular interface \cite{linear}.
Owing to its two-dimensionality, the problem can be reformulated as
a nonlocal nonlinear partial differential equation for a conformal
map which is analytic in the exterior of the unit circle
\cite{Constantin2}. This equation, however, is hard for analysis. We
consider here a simple but non-trivial case that can be analyzed
directly in the physical plane: the dynamics of a half-infinite (or,
physically, very long) stripe.

\begin{figure}
\includegraphics[width=4.5 cm,clip=]{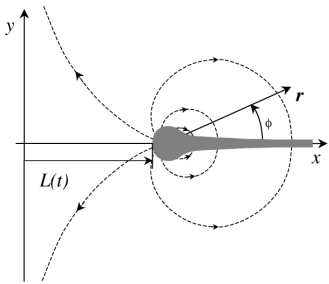}
\caption{Setting for the dumbbell dynamics.} \label{figstrip}
\end{figure}

\textit{Stripe dynamics: theoretical predictions.} Let at $t=0$ the
bubble have the form of a half-infinite straight stripe of width
$\Delta$, located along the $x$-axis as shown in
Fig.~\ref{figstrip}. The external boundary of the system $\Gamma$ is
at infinity, where the pressure is bounded. We will measure the
distance in units of $\Delta$, the time in units of $\tau=48 \mu
\Delta^3/(\pi \sigma b^2)$, and the pressure in units of $p_0=\pi
\sigma/(4\Delta)$. In the rescaled variables Eqs. (\ref{speed}) and
(\ref{jump2}) become $v_n = - \nabla_n p$ and $p\,|_{\gamma} = {\cal
K}$, so the rescaled problem is parameter-free.

\begin{figure}
\includegraphics[width=5.5 cm,clip=]{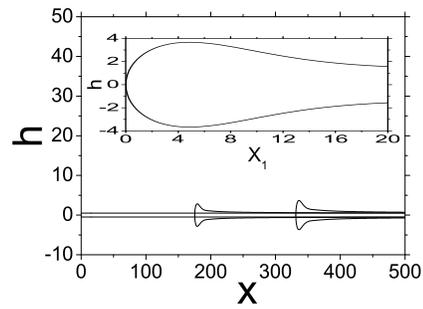}
\caption{Snapshots of a part of the simulated system at $t=0$,
$1000$ and $3010$. Notice the large difference between the
horizontal and vertical scales. The inset shows, to scale, the
lobe region at $t=3010$.} \label{figdumbbell}
\end{figure}

We are interested in the late-time behavior: $t \gg 1$. Because of
the Gibbs-Thomson effect, the pressure gradient is largest near the
tip, so the tip retreats along the x-axis. As the bubble area must
be conserved, the retreating stripe acquires a dumbbell shape, and
the lobe of the dumbbell expands with time, see
Fig.~\ref{figdumbbell}. Surprisingly, the main contribution to the
dumbbell area comes, at $t \gg 1$, from the dumbbell neck, and not
from the lobe.

Going over to a quantitative analysis, we assume (and later verify
numerically) that the lobe can be characterized by a
\textit{single} time-dependent length scale $R(t)$. Another length
scale is $L (t)$: the retreat distance of the dumbbell. Our main
objective is to find the exponents of the power laws for $R(t)$
and $L(t)$. Our analysis will \textit{not} give the numerical
coefficients of these power laws (which, in the rescaled units,
are of order unity); these will be found numerically. Introduce
polar coordinates $r$ and $\phi$, see Fig.~\ref{figstrip}. The
dumbbell neck, $r \gg R(t)$, is almost flat, so $p$ must vanish at
$\phi \to 0$ and $\phi \to 2\pi$. On the other hand, $p={\cal K}
\sim 1/R(t)$ at the lobe interface (for definiteness, at $\phi=
\pm \,\pi/2$).  Therefore, the leading term in the multipole
expansion \cite{Jackson} is
\begin{equation}
p (r, \phi, t) = C \left[R(t)\,r\right]^{-1/2}\, \sin(\phi/2) ,
\label{B1}
\end{equation}
where $C ={\cal O}(1)$. Having demanded the Gibbs-Thomson condition
here, we somewhat stretched the validity of Eq. (\ref{B1}), but this
can only affect the value of constant $C$. The dashed lines in Fig.
\ref{figstrip} show the field lines of $\nabla p$.

Equation (\ref{B1}) yields the normal component of the interface
speed $v_n = - \nabla_n p$ in the neck region. For the upper
interface of the neck
\begin{equation}
\label{v_n} v_n = -\frac{1}{r} \frac{\partial p}{\partial
\phi}(\phi\to 0) = -\frac{C}{2 R^{1/2}(t)\, r^{3/2}}\,.
\end{equation}
Now we return to the Cartesian coordinates. Let $h(x_1,t)$ be the
local height of the dumbbell, while $x_1=x-L(t)$ be the horizontal
coordinate in the moving frame with the origin at the tip. In the
neck region, $x_1 \gg R(t)$, the quantity $\partial
h(x_1,t)/\partial t$ is given by Eq. (\ref{v_n}), so we obtain
\begin{equation}\label{integral}
h(x_1,t) -\frac{1}{2} = \int_0^t \frac{C\,dt^\prime}{2
R^{1/2}(t^\prime)\, x_1^{3/2}}\,.
\end{equation}
This equation yields $h(x_1,t)$ in two different limits. At very
large distances from the lobe, $x \gg L(t)$ (region I)
\begin{equation}\label{veryfar}
   h(x_1,t) -\frac{1}{2}\simeq \frac{C}{2\,
   x_1^{3/2}}\int_0^t\frac{dt^\prime}{R^{1/2}(t^\prime)}\, \sim
   \frac{t}{x_1^{3/2}\,R^{1/2}(t)}\,,
\end{equation}
where the last estimate assumes that $R(t)$ is a power of $t$.
Another limit corresponds to intermediate distances: $R(t) \ll x_1
\ll L(t)$ (region II). Here, at fixed $x$, the main contribution
to the integral in Eq.~(\ref{integral}) comes from times close to
$t$, so that $x_1(t)/\dot{L}(t) \ll t-t^{\prime} \ll t\,$. Indeed,
one can expand $x_1(t^\prime) = x_1(t)+\dot{L}(t) (t-t^\prime)+
\dots\,$ and, in the leading order, ignore higher order terms. The
effective time interval for the integration is $(t-\delta
t^{\prime}, t)$, where $\delta t^{\prime} \sim x_1(t)/\dot{L}(t)$.
Furthermore, $R^{1/2}(t^\prime)$ can be evaluated at
$t^{\prime}=t$, as its variation  on the time interval $(t-\delta
t^{\prime}, t)$ is negligible. Then, extending the lower limit of
the integral to $-\infty$ and calculating the remaining elementary
integral, we obtain
\begin{equation}\label{far}
h(x_1,t) -\frac{1}{2}\simeq
\frac{C}{R^{1/2}(t)\,\dot{L}(t)x_1^{1/2}(t)}\,.
\end{equation}
Now we can estimate the contributions of regions I and II to the
dumbbell area gain $A$ in the neck region. We integrate Eq.
(\ref{veryfar}) over $x_1$ from, say, $2 L(t)$ to infinity, and
Eq. (\ref{far}) from $R(t)$ to $2 L(t)$. The results are:
\begin{equation}\label{regionI}
 A_I (t) \sim \frac{t}{L^{1/2}(t)\,R^{1/2}(t)}\;\;\;\;\;\;\mbox{in region I}\,,
\end{equation}
and
\begin{equation}\label{regionII}
A_{II} (t) \sim
\frac{L^{1/2}(t)}{\dot{L}(t)\,R^{1/2}(t)}\;\;\;\;\;\;\mbox{in
region II}\,.
\end{equation}
Once $L(t)$ is a power law, $A_I$ and $A_{II}$ are comparable.
Notice that in region I (respectively, II) the main contribution
comes from the lower (respectively, upper) limit of integration.
As we verify \textit{a posteriori}, the contribution to the
dumbbell area of the lobe itself, $A_R\sim R^2(t)$, is negligible
compared to $A_I$ and $A_{II}$ as long as $t \gg 1$.

Now we can find the dynamic exponents of $L(t)$ and $R(t)$. First,
we employ the area conservation of the dumbbell. The area loss
$L(t) \times 1$ of the retreating dumbbell must be equal to the
area gain in the neck, so up to numerical coefficients of order
unity
\begin{equation}
L(t)\sim A_I(t) \sim A_{II}(t) \sim
\frac{t}{L^{1/2}(t)\,R^{1/2}(t)}\,. \label{conservation}
\end{equation}
Second, there is a simple kinematic relation between $\dot{L}(t)$
and the characteristic speed of the lobe motion $V_{l}$. Using
Eq.~(\ref{B1}), we obtain  $V_{l} \sim - \partial p/\partial r
\left(r \sim R(t), \phi \simeq \pi\right) \sim R^{-2}(t)$, and
demand $\dot{L}(t)\sim R^{-2}(t)$. Combined with
Eq.~(\ref{conservation}), this yields
\begin{equation}
L(t)\sim t^{3/5} \;\;\;\mbox{and} \;\;\;R(t) \sim
 t^{1/5}\;\;\; \mbox{at}\; t \gg 1\,. \label{scalings}
\end{equation}
Now we can return to Eqs. (\ref{B1})-(\ref{regionII}) and find the
explicit time-dependences. For example, the far-neck asymptote in
Eq. (\ref{veryfar}) becomes $h(x,t)-1/2 \sim t^{9/10} x^{-3/2}$.
We can also verify that, at $t \gg 1$, the lobe area $A_R \sim
R^2(t) \sim t^{2/5}$ is indeed much less than $A_I(t)\sim
A_{II}(t)\sim t^{3/5}$.

That the lobe is characterized by a single dynamic length scale
$R(t) \sim t^{1/5}$ implies a similarity Ansatz for the lobe shape
in the moving frame:
\begin{equation}\label{Ansatz}
h_s(x_1,t)=t^{1/5} \Phi (x_1/t^{1/5})\;\;\; \mbox{at}\; t \gg 1\,.
\end{equation}

\textit{Numerical method.} To test our predictions, we performed 
simulations of the dynamics of long stripes with
dimensions $X \times 1$, where $X \gg 1$. The ultimate shape of such
a stripe is a perfect circle. Therefore, the scaling behavior,
predicted by our theory of a one-sided dumbbell, appears as an
\textit{intermediate asymptote}, 
as we require $R(t) \gg 1$ but $L(t)\ll X$. In view of the predicted
scalings with time, we must demand $1 \ll t \ll X^{5/3}$.

Our numerical algorithm \cite{VM} is based on a representation of
the harmonic field in terms of a line integral over the bounding
contour; it involves tracking of the contour nodes. We employed a
variant of the boundary integral method, suggested in Ref.
\cite{GGM}. The algorithm includes solving an integral equation for
an effective density of dipole moments (DMD) and evaluating another
integral, which yields a harmonic conjugated function (HCF). The
normal velocity of the interface is given by the derivative of the
HCF along the contour. The very large aspect ratio of the dumbbell
demands a different discretization compared to Ref. \cite{GGM}.
Indeed, the typical scale of variation of the kernel of the integral
equation \cite{GGM} over the almost flat neck of the dumbbell is
close to $1$: the initial stripe thickness. On the other hand, the
DMD changes much slower there. This enabled us to considerably
reduce the number of grid nodes in the neck region. We used a
piecewise constant function to approximate the DMD, and a piecewise
linear function to approximate the contour. Therefore, each of the
integrals was approximated as a discrete sum of the DMD values
multiplied by an integral of the kernel between two neighboring
nodes. The latter integrals can be calculated analytically. The HCF
is evaluated at middle points between the nodes, while the normal
velocity at each node is evaluated using the values of the HCF at
neighboring middle points.

We used an explicit finite difference method to track the contour.
The number of grid points, needed for an accurate solution and
contour tracking, decreases with time together with the perimeter
of the dumbbell. An obvious modification of the algorithm of Ref.
\cite{GGM} exploited the 4-fold symmetry of the dumbbell. The area
conservation of the dumbbell was used for accuracy control. The
time step chosen was $5 \times 10^{-3} \, min \,|R^i/v_n^i|$,
where $R_i$ and $v_n^i$ are the local curvature radius and normal
velocity, respectively, in the node $i$  of the contour. This
choice resulted in good area conservation: in the simulation
described below less than $0.5\%$ of the area was lost by the time
$t=7 000$. As the dumbbell contour becomes smoother, the time step
greatly increases.

\begin{figure}
\includegraphics[width=8.0 cm,clip=]{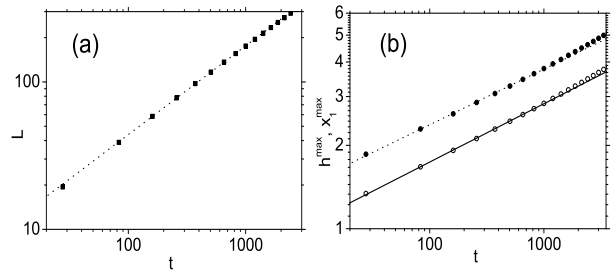}
\caption{Figure a shows, in a log-log scale, the retreat distance
$L$ versus time and its power-law fit $2.75 \,t^{0.60}$. Figure b
shows, in a log-log scale, the maximum dumbbell height, $h^{max}$
(the empty circles), and the position of the maximum, $x_1^{max}$
(the filled circles), versus time, as well as their power-law fits
$0.66\,t^{0.21}$ and $0.94\,t^{0.20}$, respectively.}
\label{lengths}
\end{figure}

\textit{Numerical results.} Here we report a simulation with
$X=2000$. Figure~\ref{figdumbbell} shows snapshots of a part of the
system at times $t=0$, $1000$ and $3010$. The stripe develops a
dumbbell shape (though some may prefer a comparison with daisy
petal). The lobe grows with time, the neck widens. Shown in
Fig.~\ref{lengths}a is the retreat distance $L(t)$ versus time. A
power law fit yields exponent 0.60 which coincides with the
theoretical value $3/5$. Figure~\ref{lengths}b shows the maximum
dumbbell height, $h^{max}$, and the position of the maximum,
$x_1^{max}$, versus time. Each of these two quantities exhibits a
power law; the fitted exponents are $0.21$ (for $h^{max}$) and
$0.20$ (for $x_1^{max}$), in agreement with the theoretical value
$1/5$. At long times, when the aspect ratio of the dumbbell is
already not large enough, the straight line in Fig.~\ref{lengths}a
slightly curves down, while those in Fig.~\ref{lengths}b curve up.
We verified that for a shorter stripe, $X=1000$, deviations from the
\textit{same} straight lines occur earlier, as expected. The time
interval of the three fits, $20<t<1000$, corresponds to the common
parts of the dependences for the two values of $X$.

\begin{figure}
\includegraphics[width=4.9 cm,clip=]{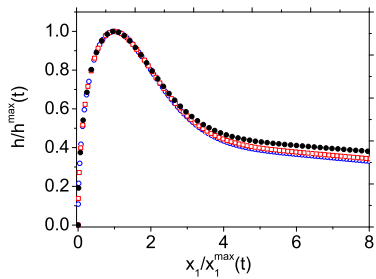}
\caption{Self-similarity of the lobe. Shown is the shape function
$h(x_1,t)$, rescaled to the maximum dumbbell elevation, versus the
coordinate $x_1$, rescaled to the abscissa of the maximum, at
times $160.3$ (the filled circles), $1000$ (the squares), and
$3010$ (the empty circles).} \label{SS}
\end{figure}

\begin{figure}[ht]
\includegraphics[width=8.5 cm,clip=]{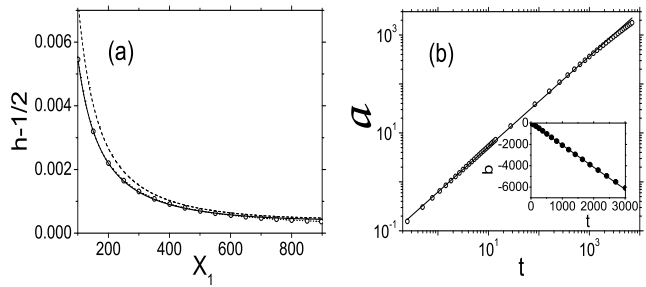}
\caption{The dumbbell neck shape and dynamics. Figure a: the neck
shape at $t=13.9$, found numerically (the dotted line with circles),
Eq.~(\ref{tailfull}) (the solid line), and the first term of
Eq.~(\ref{tailfull}) (the dashed line). Figure b and its inset show,
in a log-log and linear scales, respectively, the pre-factors $a$
and $b$ versus time (symbols). Also shown are a power-law fit with
exponent $0.92$ (figure b) and a linear fit (inset).} \label{tail}
\end{figure}

Figure~\ref{SS} depicts the (rescaled) dumbbell shape in the
moving frame at three different times. 
The collapse of three different curves into a single one supports
the similarity Ansatz (\ref{Ansatz}). Notice that the dumbbell
shape $h(x_1,t)$ in region II [$R(t)\ll x_1 \ll L(t)$] belongs to
the similarity region. Indeed, assuming that $\Phi(\xi) \sim
\xi^{-1/2}$ at $\xi \gg 1$, we see that Eq. (\ref{Ansatz}) yields
Eq. (\ref{far}) (where one should substitute $L \sim t^{3/5}$ and
$R \sim t^{1/5}$, and neglect $1/2$ in the left hand side).

The self-similarity breaks down at a distance $x_1 \sim L(t) \sim
t^{3/5}$ from the tip.  Beyond this distance, Eq.~(\ref{veryfar})
predicts a power-law neck shape. Figure~\ref{tail}a shows the
shape of the dumbbell neck at time $t=13.9$,  computed
numerically. Also shown are the quantity
\begin{eqnarray}\label{tailfull}
h(x_1)-\frac{1}{2}=a\, \left[x_1^{-3/2}+(X-2 L-x_1)^{-3/2}\right]
+ \nonumber
\\
b\,\left[x_1^{-2}+(X-2 L-x_1)^{-2}\right]\,,
\end{eqnarray}
and its first term, proportional to $a$.  Equation (\ref{tailfull})
differs from Eq. (\ref{veryfar}) in that (i) it accounts for
contributions from \textit{two} dumbbell lobes, and (ii) it accounts
for the sub-leading term in the multipole expansion of the harmonic
function $p$. Note that $a$ and $b$ are the only adjustable
parameters in Eq. (\ref{tailfull}). The resulting profile is almost
indistinguishable from the numerical profile. The first term of Eq.
(\ref{tailfull}) already gives fairly good agreement. The excellent
agreement holds, on a shrinking interval of $x_1$, until $t=7020$.
Figure~\ref{tail}b and its inset show $a$ and $b$ versus time,
respectively. A power-law fit of $a(t)$ yields exponent $0.92$,
close to our prediction $9/10$. The pre-factor $b(t)$ behaves
linearly with time. How does it compare with the theory? As $p \sim
1/R(t)$ at the lobe interface, the sub-leading term $p \sim r^{-1}
\sin \phi$ does not include $R(t)$. Then, repeating the procedure
which led us to Eq.~(\ref{veryfar}), we do obtain $b(t) \sim t$.

\textit{Summary.} We studied the UHS flow in the case when the
inviscid fluid is initially in the form of a long stripe. We found
that the resulting dumbbell dynamics exhibit self-similarity with
nontrivial exponents. The solution we obtained is the first
analytical solution for an UHS flow that goes beyond a linear
theory. Similarly to other curve-shortening area-preserving
relaxation models \cite{other}, the stripe relaxation provides a
useful characterization of this non-integrable flow. Its
experimental realization should not be difficult.

We thank Eran Sharon for a useful discussion. This work was
supported by the Israel Science Foundation (Grant No. 180/02), and
by the Russian Foundation for Basic Research (Grant No.
05-01-000964).


\begin{thebibliography}{111}
\bibitem{linear} The damping rates of small sinusoidal
perturbations of flat and circular interfaces are given by the
zero-flow-rate limit of Eq.~(10) of Ref. \cite{ST} (flat
interface), and of Eq.~(11) of Ref. \cite{Paterson} (circular
interface).
\end{thebibliography}
\end{document}